\documentclass{ws-procs9x6}

\begin{document}

\title{Neutrino properties from high energy astrophysical neutrinos}

\author{Sandip Pakvasa\address{Department of Physics and Astronomy, University of Hawaii, 
        Honolulu, HI  96822,  USA} \\}

\maketitle

\abstracts{
It is shown how high energy neutrino beams from 
very distant sources can be utilized to learn about some properties
of neutrinos such as lifetimes, mass hierarchy, etc.  Furthermore, 
even mixing elements such as $U_{e3}$ and the CPV phase in the 
neutrino mixing matrix can be measured
in principle.  Pseudo-Dirac mass differences as small
as $10^{-18} eV^2$ can be probed as well.}

\section{Introduction}
We make two basic assumptions which are reasonable.  The first one is that distant neutrino
sources (e.g. AGN's and GRB's) exist; and furthermore with detectable fluxes
at high energies (upto and beyond PeV).  The second one is that in the not
too far future, very large volume, well instrumented detectors of sizes of
order of KM3 and beyond will exist and be operating; and furthermore will
have (a) reasonably good energy resolution and (b) good angular resolution
($\sim 1^0 $ for muons).

\section{Neutrinos from Astrophysical Sources}

If these two assumptions are valid, then there are a number of uses these
detectors can be put to\cite{pakvasa}.  In this talk I want to focus on those that enable us
to determine some properties of neutrinos: namely, probe neutrino lifetimes 
to $10^4 s/eV$ (an improvement of $10^8$ over current bounds), 
pseudo-Dirac mass splittings to a level of $10^{-18} eV^2$ (an improvement of a factor of $10^6$ over
current bounds) and potentially even measure quantities such as $U_{e3}$ and the phase $\delta$ in the MNSP matrix\cite{MNS}.

\section{Astrophysical neutrino flavor content}

In the absence of neutrino oscillations we expect a very small $\nu_\tau$
component
in neutrinos from astrophysical sources. From the most discussed and  the most likely astrophysical high energy
neutrino sources\cite{learned} we expect nearly equal numbers of particles
and anti-particles, half as many $\nu_e's$ as $\nu_\mu's$ and virtually no $\nu_\tau's$.  This comes about simply 
because the neutrinos are thought to originate in decays of pions (and
kaons) and subsequent decays of muons.  Most astrophysical targets are fairly tenous even compared to the Earth's atmosphere, and would allow for full muon decay  
in flight.  There are some predictions for flavor independent fluxes from cosmic defects and exotic objects such as evaporating black holes.  Observation of
tau neutrinos from these would have great importance.  A conservative estimate\cite{learned1} shows that the prompt $\nu_\tau$ flux is very small and the emitted flux is close to the ratio $1:2:0$.  The flux ratio of $\nu_e: \nu_\mu: \nu_\tau = 1:2:0$
is certainly valid for those AGN models in which the neutrinos are 
produced in beam dumps of photons or protons on matter, in which
mostly pion and kaon decay(followed by the decay of muons) supply the bulk of 
the neutrino flux. 

Depending on the amount of prompt $\nu-$flux due to the production and decay
of heavy flavors, there could be a small non-zero $\nu_\tau$ component 
present. 

\section{Effect of Oscillations}

The current knowledge of neutrino masses and mixings can be summarized 
as follows\cite{pakvasa1}. The mixing matrix elements are given to a good approximation with 
the solar mixing angle
given by about $32^0$, the atmospheric angle by about $45^0$ and $U_{e3} < 0.17$ limited by the CHOOZ bound.  The mass spectrum has two possibilities; normal or inverted, and 
with the mass differences given by $\delta m^2_{32} \sim 2.10^{-3} eV^2$ and 
$\delta m_{21}^2 \sim 7.10^{-5} eV^2$.  Since $\delta m^2 L/4E$ for 
the distances to GRB's and AGN's (even for energies upto and beyond PeV) 
is very large $(> 10^7)$ the oscillations have always averaged out 
and the conversion(or survival) probability is given by
\begin{eqnarray}
P_{\alpha \beta} &=& \sum_{i} | U_{\alpha i} \mid^2 \mid U_{\beta i} \mid^2 \
\end{eqnarray}
Assuming no significant matter effects enroute, it is easy to show that
the mixing matrix in Eq. (1) leads to a propagation matrix P, 
which, for any value of the solar mixing angle, converts a flux ratio of 
$\nu_e: \nu_\mu: \nu_\tau = 1:2:0$ 
into one of $1:1:1$.  Hence the flavor mix expected at arrival 
is simply an equal mixture of $\nu_e, \nu_\mu$ and 
$\nu_\tau$ as was observed long ago\cite{learned1,athar}.
If this universal flavor mix is confirmed by future observations, our current
knowledge of neutrino masses and mixings is reinforced and conventional
wisdom about the beam dump nature of the production process is confirmed as
well. However, it would much more exciting to find deviations from it, and learn something new. How can this come about? Below is a shopping list of a variety of ways in which this could come to pass.
  
\section{Deviations from Canonical Flavor Mix}

There are quite a few ways in which the flavor mix can be changed from
the simple universal mix. 

The first and simplest is that initial flavor mix is NOT $1:2:0$. This
can happen when there are strong magnetic fields causing muons to lose
energy before they decay, and there exist models for neutrino production
in AGN's in which this does happen\cite{rachen}. In this case the $\nu_e's$ have much 
lower energies
compared to $\nu_\mu's$ and effectively the initial flavor mix is
$0:1:0$ and averaged out oscillations convert this into  $1/2:1:1$
on arrival.

The possibility that the mass differences between neutrino mass
eigenstates are zero in vacuum (and become non-zero only in the presence
of matter) has been raised recently\cite{kaplan}. If this is true, then the final flavor mix 
should be the same as initial, namely: $1:2:0.$

  Neutrino decay is another important possible way for the flavor mix to 
deviate significantly from the democratic mix\cite{beacom}.We now know that neutrinos have non-zero masses and non-trivial mixings,
based on the evidence for neutrino mixings and oscillations from the data on
atmospheric, solar and reactor neutrinos.

If this is true, then in general, the heavier neutrinos are expected to 
decay into the lighter ones via flavor changing processes\cite{pakvasa2}.  
The only questions are (a) whether the lifetimes are 
short enough to be phenomenologically interesting (or are 
they too long?) and (b) what are the dominant decay modes.

Since we are interested
in decay modes which are likely to have rates (or lead to lifetimes)  which
are phenomenologically interesting,  we can rule out several classes of decay
modes immediately. For example, the very strong constraints on radiative decay
modes and on three body modes such as $\nu \rightarrow 3\nu$ render them
as being uninteresting.

The only decay modes 
which can have interestingly fast decays rates are two body 
modes such as $\nu_i \rightarrow \nu_j + x$  where 
$x$ is a very light or massless particle, e.g. a Majoron.
In general, the Majoron is a mixture of the Gelmini-Roncadelli\cite{gelmini} and Chikasige-Mohapatra-Peccei\cite{chicasige} type Majorons.  The effective interaction is of the form:
\begin{equation}
\bar{\nu}^c_\beta (a+b \gamma_5) \nu_\alpha \ x
\end{equation}
giving rise to decay:
\begin{equation}
\nu_\alpha \rightarrow \bar{\nu}_\beta \ ( or  \ \nu_\beta)  +  x
\end{equation}

where $x$ is a massless, spinless particle; $\nu_\alpha$ and $\nu_\beta$
are mass eigenstates which may be mixtures of flavor and sterile neutrinos.
Explicit models of this kind which can give rise to fast neutrino decays have
been discussed\cite{valle}.
These models are unconstrained by $\mu$ and $\tau$ decays which do not arise
due to the $\Delta L = 2$ nature of the coupling.
 The couplings of $\nu_\mu$ and $\nu_e$ are constrained
by the limits on multi-body $\pi$, K decays,
and on $\mu-e$ university violation in $\pi$ and K decays\cite{barger}, 
but these bounds allow fast neutrino decays.

There are very interesting cosmological implications of such couplings. The
details depend on the spectrum of neutrinos and the scalars in the model.
For example, if all the neutrinos are heavier than the scalar; the relic
neutrino density vanishes today, and the neutrino mass bounds from CMB
and large scale structure are no longer operative, whereas future
measurements in the laboratory might find a non-zero result for a neutrino
mass \cite{beacom1}. 
If the scalars are heavier than the neutrinos, there are signatures
such as shifts of the $n$th multipole peak (for large $n)$ in the 
CMB \cite{chacko}.
There are other implications as well, such as the number of relativistic
degrees of freedom(or effective number of neutrinos) being different at the
BBN and the CMB eras. The additional degrees of freedom should be detectable
in future CMB measurements.

Direct limits on such decay modes are also very weak.
Current bounds on such decay modes are as follows.  For the mass eigenstate $\nu_1$, the limit is about
\begin{equation}
\tau_1 \geq 10^5 \ sec /eV
\end{equation}
based on observation of $\bar{\nu}_e's$ from SN1987A \cite{hirata}
(assuming CPT invariance). For $\nu_2$, strong  limits can be deduced  from
the non-observation of solar anti-neutrinos in KamLAND\cite{eguchi}, 
a more general bound is obtained from 
an analysis of solar neutrino data\cite{bell} leads to a bound given by:
\begin{equation}
\tau_2 \geq 10^{-4} \ sec/eV
\end{equation}
For $\nu_3$, in case of  normal hierarchy, one can derive a bound from the atmospheric neutrino observations of upcoming neutrinos\cite{barger1}:
\begin{equation}
\tau_3 \geq \ 10^{-10} \ sec/eV
\end{equation}

The strongest lifetime limit is thus too weak to eliminate the possibility of
astrophysical neutrino decay by a factor about $10^7 \times (L/100$ Mpc) 
$\times (10$ TeV/E).  It was noted that the
disappearance of all states except $\nu_1$ would prepare a beam that could in principle be used to measure elements of the neutrino mixing matrix, namely the ratios $U^2_{e1} : U^2_{\mu 1} : U^2_{\tau 1}$\cite{pakvasa3}.  
The possibility of measuring
neutrino lifetimes over long baselines was mentioned in Ref.\cite{weiler}, 
and some predictions for decay in four-neutrino models were given in 
Ref.\cite{keranen}.  The particular values and small 
uncertainties on the neutrino mixing parameters allow 
for the first time very distinctive signatures of the effects of 
neutrino decay on the detected flavor ratios.  
The expected increase in neutrino lifetime sensitivity (and corresponding 
anomalous 
neutrino couplings) by several orders of magnitude makes for a very
interesting test of physics beyond the Standard Model; a discovery would
mean physics much more exotic than neutrino mass and mixing alone.   
Neutrino decay because of its unique signature cannot be mimicked by either different neutrino flavor ratios at the source or other non-standard neutrino interactions.

A characteristic feature of decay is its strong energy dependence: 
$\exp (-Lm/E \tau)$, where $\tau$ is the rest-frame lifetime.  
For simplicity, consider the case that decays are always complete, i.e., that 
these exponential factors vanish.  


The simplest case (and the most generic expectation) is a normal hierarchy 
in which both $\nu_3$ and $\nu_2$ decay, leaving only the 
lightest stable eigenstate  $\nu_1$.  In this case the 
flavor ratio is 
$U^2_{e1}:  U^2_{\mu 1} : U^2_{\tau 1}$\cite{pakvasa3}. 
Thus if $U_{e3} = 0$
\begin{equation}
\phi_{\nu e} :  \phi_{\nu_{\mu}} :  \phi_{\nu_{\tau}}
\simeq 5 : 1 : 1, 
\end{equation}
where we used the neutrino mixing parameters given above\cite{beacom}.  
Note that this is an extreme deviation of the flavor ratio from
that in the absence of decays.  It is difficult to imagine other mechanisms
that would lead to such a high ratio of $\nu_e$ to $\nu_\mu$.  In the case
of inverted hierarchy, $\nu_3$ is the lightest and hence stable state, and
so\cite{beacom}
\begin{equation}
\phi_{\nu_{e}} :  \phi_{\nu_{\mu}} : \phi_{\nu _{\tau}} = U^2_{e3} : 
U^2_{\mu 3} : U^2_{\tau 3} = 0 : 1 : 1.
\end{equation}
If  $U_{e3} = 0$ and $\theta_{atm} = 45^0$, each mass eigenstate has equal
$\nu_\mu$ and $\nu_\tau$ components.  Therefore, decay cannot break 
the equality between the $\phi_{\nu_{\mu}}$ and $\phi_{\nu_{\tau}}$ 
fluxes and thus the $\phi_{\nu_{e}} : \phi_{\nu_\mu}$ ratio contains all the useful information. 
The effect of a non-zero $U_{e3}$ on the no-decay case of 1 : 1 : 1 
is negligible.

When $U_{e3}$ is not zero, and the hierarchy is normal, it is possible to
obtain information on the values of $U_{e3}$ as well as the CPV phase $\delta$\cite{beacom2}.  The flavor ratio $e/\mu$ varies from 5 to 15 (as $U_{e3}$ goes from 0 to 0.2)  
for $\cos \delta =+1$ but from 5 to 3 for $\cos \delta =-1$.  The ratio $\tau/\mu$ varies from 1 to 5 $(\cos \delta = +1)$ or 1 to 0.2 $(\cos \delta =-1)$ for the same range of $U_{e3}$.

If the decays are not complete and if the daughter does not carry the full
energy of the parent neutrino; the resulting flavor mix is somewhat
different but any case it is still quite distinct from the simple $1:1:1$
mix\cite{beacom}.



If the path of neutrinos takes them thru regions with significant magnetic 
fields and the neutrino magnetic moments are large enough, the flavor mix can 
be affected\cite{enquist}.  The main effect of the passage thru magnetic field is the 
conversion of a given helicity into an equal mixture of both helicity states.
This is also true in passage thru random magnetic fields\cite{domokos}.

If the neutrino are Dirac particles, and all magnetic moments are comparable, 
then the effect of the spin-flip is to simply reduce the overall flux of all 
flavors by half, the other half becoming the sterile Dirac partners.
If the neutrinos are Majorana particles, 
the flavor composition remains 1 : 1 : 1 when it 
starts from 1 : 1 : 1, and the absolute flux remains unchanged.

What happens when large magnetic fields are present in or near the neutrino 
production region?  In case of Dirac neutrinos, there is no difference and 
the outcoming flavor ratio remains 1 : 1 : 1, with the absolute fluxes
reduced by half.  In case of Majorana neutrinos,
since the initial flavor mix is no longer universal but is 
$\nu_e: \nu_\mu: \nu_\tau \approx 1: 2: 0,$ this is modified
but it turns out that the final(post-oscillation) flavor mix is still 1 : 1
: 1  !

Other neutrino properties can also affect the neutrino flavor mix and modify
it from the canonical 1 : 1 : 1. If neutrinos have flavor(and equivalence
principle) violating couplings to gravity(FVG), or Lorentz invariance
violating couplings; then there can be resonance
effects which make for one way transitions(analogues of MSW transitions)
e.g. $\nu_\mu \rightarrow \nu_\tau$ but not vice
versa\cite{minakata,barger3}. In case of FVG for example,
this can give rise to an anisotropic deviation of the $\nu_\mu/\nu_\tau$
ratio from 1, becoming less than 1 for events coming from the direction
towards the Great
Attractor, while remaining 1 in other directions\cite{minakata}. 
 
Another  possibility
that can give rise to deviations of the flavor mix from the canonical
1 : 1 : 1 is the idea of neutrinos of varying mass(MaVaNs). In this
proposal\cite{fardon}, by having the dark energy and neutrinos(a sterile one to be
specific) couple, and track each other; it is possible to relate the
small scale $2\times 10^{-3}$ eV required for the dark energy to the small
neutrino mass, and furthermore the neutrino mass depends inversely
on neutrino density, and hence on the epoch. As a result, if this
sterile neutrino mixes with a flavor neutrino, the mass difference
varies along the path, with potential resonance enhancement of the
transition probability into the sterile neutrino, and thus change the
flavor mix\cite{hung}. For example, if only one 
resonance is crossed enroute, it can 
lead to a conversion of the lightest (mostly) 
flavor state into the (mostly) sterile
state, thus changing the flavor mix to
 $1-U_{e1} ^2 \ : 1-U_{\mu 1}^2 \ : 
1-U_{\tau 1}^2
\approx 1/3 \ : 1 \ : 1,$ in case of inverted 
hierarchy and similarly $\approx 2 \ : 1 \ : 1$
in case of normal hierarchy.

Complete quantum decoherence would give rise to a flavor mix given
by $1:1:1$, which is identical to the case of averaged out oscillations
as we saw above. The distinction is that complete decoherence always
leads to this result; whereas averaged out oscillations lead to this
result only in the special case of the initial flavor mix being $1:2:0.$
To find evidence for decoherence, therefore, requires a source which
has a different flavor mix . One possible practical example is a source
which emits $\nu_e's$ by decay of neutrons, and hence no $\nu_\mu's$
at all, with an initial flavor mix of $1:0:0$. In this case  decoherence 
leads to the universal $1:1:1$ mix whereas the averaged out oscillations
lead to $3:1:1$\cite{hooper}. The two cases can be easily distinguished from each other.


If each of the three neutrino mass eigenstates is actually a doublet 
with very small mass difference (smaller than $10^{-6} eV)$, 
then there are no current experiments  that could have detected this. 
Such a possibility was raised long ago\cite{bilenky1}. 
It turns out that the only way to detect such small mass 
differences $(10^{-12} eV^2 > \delta m^2 > 10^{-18} eV^2)$ 
is by measuring flavor mixes of the high energy neutrinos 
from cosmic sources.  Relic supernova neutrino
signals and AGN neutrinos are sensitive to mass difference 
squared down to $10^{-20} eV^2$ \cite{beacom3}.

Let $(\nu_1^+, \nu_2^+, \nu_3^+; \nu_1^-,  \nu_2^-, \nu_3^-)$ 
denote the six mass eigenstates where $\nu^+$ and $\nu^-$ are a 
nearly degenerate pair.  A 6x6 mixing matrix rotates the mass 
basis into the flavor basis.  
In general, for six Majorana neutrinos, there would be fifteen 
rotation angles and fifteen phases.  However, for pseudo-Dirac 
neutrinos, Kobayashi and Lim\cite{kobayashi} have given an elegant proof 
that the 6x6 matrix $V_{KL}$ takes the very simple form 
(to lowest order in $\delta m^2 / m^2$:
\begin{eqnarray}
V_{KL} = \left (
\begin{array}{cc}
U & 0 \\
0 & U_R 
\end{array} \right) \cdot
\left (
\begin{array}{cc}
V_1 & iV_1 \\
V_2  & -iV_2 
\end{array}\right),
\end{eqnarray}
where the $3\times 3$ matrix U is just the usual mixing 
matrix determined by the atmospheric and solar observations, the 
$3\times 3$ matrix $U_R$ is an unknown unitary matrix and $V_1$ and 
$V_2$ are the diagonal matrices 
$V_1 =$ diag $(1,1,1)/\sqrt{2}$, and $V_2$=diag$(e^{-i \phi_1}, 
e^{-i \phi_2}, e^{-i \phi_3})/\sqrt{2}$, with the $\phi_i$ 
being arbitrary phases.


As a result, the three active neutrino states are described in terms of the six mass eigenstates as:
\begin{equation}
\nu_{\alpha L} = U_{\alpha j} \ \frac{1}{\sqrt{2}} \left(\nu^+_{j} + i \nu^-_{j}  \right).
\end{equation}

The flavors deviate from the democratic value of $\frac{1}{3}$ by
\begin{eqnarray*}
\label{deviate3}
\delta P_e &=& -\frac{1}{3}\,\left[ \frac{3}{4}\chi_1 + \frac{3}{4}\,
\chi_2 \right],\nonumber\\
\delta P_\mu = \delta P_\tau  &=& -\frac{1}{3}\,\left[ 
\frac{1}{8}\chi_1 + \frac{3}{8}\,\chi_2 
	+ \frac{1}{2}\,\chi_3\right] \,
\end{eqnarray*} 
where $\chi_i = \sin^2(\delta m_i^2 L/4E)$.The flavor ratios deviate 
from $1:1:1$ when
one or two of the pseudo-Dirac oscillation modes is accessible.  In
the ultimate limit where $L/E$ is so large that all three oscillating
factors have averaged to $\frac{1}{2}$, the flavor ratios return to $1:1:1$,
with only a net suppression of the measurable flux, by a factor of
$1/2$.

\section{\bf Cosmology with Neutrinos} 
If the oscillation phases can indeed be measured for the very small 
mass differences by the deviations of the flavor mix from 1 : 1 : 1 as 
discussed above, the
following possibility is raised. It is a fascinating fact 
that non-averaged oscillation phases,
$\delta\phi_j=\delta m_j^2 t/4p$, and hence the factors $\chi_j$, are
rich in cosmological information\cite{weiler,weiler1}.  
Integrating the phase
backwards in propagation time, with the momentum blue-shifted, 
one obtains
\begin{eqnarray}
\delta\phi_j&=&\int_0^{z_e} dz\frac{dt}{dz}\frac{\delta m_j^2}{4p_0(1+z)}\\
	  &=&\left(\frac{\delta m_j^2 H^{-1}_0}{4p_0}\right) \ I
\end{eqnarray}
where I is given by
\begin{equation}
I=\int_1^{1+{z_e}}\frac{d\omega}{\omega^2}
		\frac{1}{\sqrt{\omega^3\,\Omega_m+(1-\Omega_m)}}\,,\nonumber
\end{equation}
$z_e$ is the red-shift of the emitting source, and $H_0^{-1}$ is
the Hubble time, known to 10\%~\cite{freedman}.  
This result holds
for a flat universe, where 
$\Omega_m+\Omega_\Lambda=1$, with $\Omega_m$ and 
$\Omega_\Lambda$ the matter and vacuum energy
densities in units of the critical density.
The integral $I$
is the fraction of the Hubble time 
available for neutrino transit.
For the presently preferred values $\Omega_m=0.3$ and
$\Omega_\Lambda=0.7$, the asymptotic ($z_e\rightarrow\infty$) value of
the integral is 0.53.  This limit is approached rapidly: at
$z_e=1\,(2)$ the integral is 
77\% (91\%) saturated.  
For cosmologically distant ($z_e > 1$) sources such as gamma-ray bursts, 
non-averaged oscillation data would, in principle, allow one to
deduce $\delta m^2$ to about 20\%, without even knowing the source
red-shifts. Known values of $\Omega_m$ and $\Omega_\Lambda$ might 
allow one to infer the source redshifts $z_e$, or vice-versa.

This would be the first measurement of a cosmological
parameter with particles other than photons.  An advantage of
measuring cosmological parameters with neutrinos is the fact that
flavor mixing is a microscopic phenomena and hence presumably free of
ambiguities such as source evolution or standard candle
assumptions\cite{weiler,stodolsky}.  Another method of
measuring cosmological parameters with neutrinos is given in
Ref.\cite{choubey}.

\section{Experimental Flavor Identification} 

It is obvious from the above discussion that flavor identification is
crucial for the purpose at hand. In a water cerenkov detector flavors can be 
identified as follows.

The $\nu_\mu$ flux can be measured by the $\mu's$ produced by the charged 
current interactions and the resulting $\mu$ tracks in the detector which
are long  at these energies.  $\nu_{e}'{s}$ produce showers by both
CC and NC interactions.  The total rate for showers includes those
produced by NC interactions of $\nu_\mu's$ and $\nu_\tau's$ as well and those
have to be (and can be) subtracted off to get the real flux of $\nu_e's$.
Double-bang and lollipop events are signatures unique to 
tau neutrinos, made
possible by the fact that tau leptons decay before they lose a significant
fraction of their energy.  A double-bang event consists of a hadronic shower
initiated by a charged-current interaction of the $\nu_\tau$ followed by a
second energetic shower from the decay  of the
resulting tau lepton\cite{learned1}.  A lollipop event consists of the second
of
the double-bang showers along with the reconstructed tau lepton track (the
first bang may be detected or not).  In principle, with a sufficient number
of
events, a fairly good estimate of the flavor ratio $\nu_e: \nu_\mu: \nu_\tau$ 
can be reconstructed, as has been discussed recently.
Deviations of the flavor ratios 
from $1:1:1$ due to possible decays are so extreme that they should be
readily identifiable\cite{beacom4}. Future high energy neutrino telescopes,
such as Icecube\cite{karle}, will not have perfect ability to separately 
measure the neutrino flux in each flavor.  However, the situation is
salvagable. In the limit of $\nu_\mu - \nu_\tau$ symmetry 
the fluxes for $\nu_\mu$ and 
$\nu_\tau$ are always in the ratio 1 : 1, with or without decay. 
This is useful since the $\nu_\tau$ flux is the hardest to measure. 

Even when the  tau events are not at all 
identifiable, the relative number of shower events to track events can 
be related to the most interesting quantity for testing decay scenarios, 
i.e., the $\nu_e$ to $\nu_\mu$ ratio.  The precision of the upcoming 
experiments should be good enough to test the extreme flavor ratios produced
by decays.  If electromagnetic and hadronic  showers can be separated, then 
the precision will be even better\cite{beacom4}.Comparing, for example, the 
standard flavor ratios of 1 : 1 : 1 to the
possible 5 : 1 : 1 generated by decay, the more numerous electron neutrino
flux will result in a substantial increase in the relative number of 
shower events.The measurement will be limited by the energy resolution of the
detector and the ability to reduce the atmospheric neutrino background which 
drops rapidly with energy and 
should be negligibly small at and above the PeV scale.

\section{Discussion and Conclusions}  

The flux ratios we discuss are energy-independent because we have assumed that the ratios at production are energy-independent, that all oscillations are averaged out, and that all possible decays are complete.  In the standard scenario with only oscillations, the final flux ratios are $\phi_{\nu_{e}} :  \phi_{\nu_{\mu}} : 
 \phi_{\nu_{\tau}} = 1 : 1 : 1$.  In the cases with decay, we have found rather
different possible flux ratios, for example 5 : 1 : 1 in the normal hierarchy and 
0 : 1 : 1 in the inverted hierarchy.  These deviations from 1 : 1 : 1 
are so extreme that they should be readily measurable.

If we are very fortunate\cite{barenboin}, we may be able to observe a
reasonable number of events from several sources (of known distance) 
and/or over a sufficient range in energy.  Then the resulting 
dependence of the flux ratio $(\nu_e/\nu_\mu)$ on L/E as 
it evolves from say 5 (or 0) to 1, can 
be clear evidence of decay and further can pin down the actual lifetime
instead of just placing a bound.

To summarize, we suggest that if future measurements of the flavor mix at
earth of high energy astrophysical neutrinos find it to be
\begin{equation}
\phi_{\nu_{e}} / \phi_{\nu_{\mu}} / \phi_{\nu_{\tau}} = \alpha / 1 / 1 ;
\end{equation}
then
\begin{itemize}
\item[(i)] $\alpha \approx 1$ (the most boring case) confirms our knowledge of the
MNSP\cite{MNS} matrix and our prejudice about the production mechanism;
\item[(ii)] $\alpha \approx 1/2$ indicates that the source emits pure
$\nu_\mu's$ and the mixing is conventional;
\item[(iii)]$\alpha \approx 3$ from a unique direction, e.g. the Cygnus region, would be
evidence in favour of a pure $\bar{\nu}_e$ production as has been suggested
recently\cite{goldberg};
\item[(iv)] $\alpha > 1$ indicates that neutrinos are decaying with normal
hierarchy; and 
\item[(v)]$\alpha \ll 1$ would mean that neutrino decays are occuring with
inverted hierarchy;
\item[(vi)] Values of $\alpha$ which cover a broader range (3 to 15) and 
deviation of the $\mu/\tau$ ratio from 1(between 0.2 to 5) can yield valuable 
information about $U_{e3}$ and $\cos \delta$. Deviations of $\alpha$
which are less extreme(between 0.7 and 1.5) can also probe very small pseudo-Dirac 
$\delta m^2$ (smaller than $10^{-12} eV^2$).
\end{itemize}  

Incidentally, in the last three cases, the results have absolutely no
dependence on the initial flavor mix, and so are completely free of any
dependence on the production model. So either one learns about the production
mechanism and the initial flavor mix, as in the first three cases, or one
learns only about the neutrino properties, as in the last three cases.
To summarise, the measurement of neutrino flavor mix at neutrino telescopes
is absolutely essential to uncover new and interesting physics of neutrinos. 
In any case, it should be evident that the construction of very large neutrino detectors is a ``no lose'' proposition.

\section{Acknowledgements}
This talk is based on work in collaboration with
John Beacom, Nicole Bell, Dan Hooper, John Learned  and Tom Weiler. I thank them for a most
enjoyable collaboration. I would like to thank  the organisers of PASCOS 2004 for the
invitation to give this talk as well as their hospitality and for providing 
a most stimulating atmosphere during the meeting.
This work was supported in part by U.S.D.O.E. under grant DE-FG03-94ER40833.


\begin{thebibliography}{99}

\bibitem{pakvasa}S. Pakvasa, 9th International Symposium 
on Neutrino Telescopes, Venice, Italy, 6-9 Mar 2001, Venice 2001, Neutrino 
Telescopes, ed. M. Baldo-Ceolin, Vol. 2, p. 603; hep-ph/0105127.

\bibitem{MNS}Z. Maki, M. Nakagawa and S. Sakata, {\it Prog. Theoret. Phys.} {\bf 28}, 870 (1962); V. N. Gribov and B. M. Pontecorvo, {\it Phys. Lett.}
{\bf B28}, 493 (1969); B. W. Lee et al., {\it Phys. Rev. Lett.} {\bf 38}, 937 (1977).

\bibitem{learned}J. G. Learned and K. Mannheim, 
{\it Ann. Rev. Nucl. Part. Sci.} {\bf 50}, 603 (2000), 
and references therein.

\bibitem{learned1}J. G. Learned and S. Pakvasa, 
{\it Astropart. Phys.} {\bf 3},
267 (1995); hep-ph/9405296.

\bibitem{pakvasa1}S. Pakvasa and J. Valle,  {\it Proc. Indian. Natl. Sci. Acad.}
{\bf 70A}, 189 (2003); hep-ph/0301061.

\bibitem{athar}H. Athar, M. Jezabek and O. Yasuda, {\it Phys. Rev.} 
{\bf D62}, 103007 (2000); hep-ph/0005104; L. Bento, P. Keranen and J. Maalampi, {\it Phys. Lett.} {\bf B476}, 205 (2000); hep-ph/9912240.

\bibitem{rachen}J. P. Rachen and P. Meszaros, {\it Phys. Rev} {\bf D58}, 
123005 (1998); astro-ph/9802280.

\bibitem{kaplan}D. B. Kaplan, A. E. Nelson and N. Weiner, {\it Phys. Rev. Lett} {\bf 93} 091801 (2004).

\bibitem{beacom}J. F. Beacom, N. Bell, D. Hooper, S. Pakvasa and T.J. Weiler,
{\it Phys. Rev. Lett.} {\bf 91}, 181301 (2003); hep-ph/0211305.

\bibitem{pakvasa2}S. Pakvasa, {\it Physics Potential and Development of Muon Colliders}, Mu 99, San Francisco, AIP Conf. Proc. {\bf 542} (2000) 99, ed D. Cline, p. 99; hep-ph/0004077.





\bibitem{gelmini}G. Gelmini and M. Roncadelli, {\it Phys. Lett.} {\bf B99},
411 (1981).

\bibitem{chicasige}Y. Chikasige, R. Mohapatra and R. Peccei, {\it Phys. Rev. Lett.} {\bf 45}, 1926 (1980).

\bibitem{valle}J. Valle, {\it Phys. Lett.}, {\bf B131}, 87 (1983); G. Gelmini
and J. Valle, {\it ibid}. {\bf B142}, 181 (1983); A. Joshipura and S. Rindani, {\it Phys.
Rev.} {\bf D46}, 3008 (1992); A. Acker, A. Joshipura and S. Pakvasa, {\it Phys. Lett.} {\bf B285}, 371 (1992); A. Acker, S. Pakvasa and J. Pantaleone, {\it Phys. Rev.} {\bf D45}, 1  (1992).


\bibitem{barger}V. Barger, W-Y. Keung and S. Pakvasa, {\it Phys. Rev.} {\bf D25}, 907 (1982).

\bibitem{beacom1}J.F. Beacom, N. Bell and S. Doddelson, hep-ph/0404585. 

\bibitem{chacko} Z. Chacko, L.J. Hall, T. Okui and S.J. Oliver, hep-ph/0312267.

\bibitem{hirata} K. Hirata et al., {\it Phys. Rev. Lett.} {\bf 58}, 1497
(1988); R.M. Bionta et al., ibid. 58, 1494 (1988).

\bibitem{eguchi}K. Eguchi et al; {\it Phys. Rev. Lett.} {\bf 92}, 071301 
(2004); hep-ex/0310047.

\bibitem{bell}J. F. Beacom and N. Bell; {\it Phys. Rev.} {\bf D65}, 113009
(2002); hep-ph/0204111; and references cited therein.

\bibitem{barger1}V. D. Barger, J. G. Learned, S. Pakvasa and T. J. Weiler, {\it Phys. Rev. Lett.} {\bf 82}. 2640 (1999); hep-ph/9810121; Y. Ashie et al., hep-ex/0404034.

\bibitem{pakvasa3}S. Pakvasa, {\it Lett. Nuov. Cimm.} {\bf 31}, 497 (1981); Y. Farzan and A. Smirnov, {\it Phys. Rev.} {\bf D65}, 113001 (2002); hep-ph/0201105.

\bibitem{weiler}T. J. Weiler, W. A. Simmons, S. Pakvasa and J. G. Learned; hep-ph/9411432.

\bibitem{keranen}P. Keranen, J. Maalampi and J. T. Peltonieni, {\it Phys. Lett.} {\bf B461}, 230 (1999); hep-ph/9901403.



\bibitem{beacom2}J. F. Beacom, N. Bell, D. Hooper, S. Pakvasa and T. J.
Weiler, {\it Phys. Rev} {\bf D69}, 017303 (2004); hep-ph/0309267.



\bibitem{enquist} K. Enqvist, P. Keranen and J. Maalampi, {\it Phys. Lett.}{\bf B438},295(1998); hep-ph/9806392.

\bibitem{domokos}G. Domokos and S. Kovesi-Domokos, {\it Phys. Lett} {\bf B410}, 57 (1997); hep-ph/9703265.

\bibitem{minakata}H. Minakata and A. Yu. Smirnov, {\it Phys. Rev.} {\bf D54}, 
3698 (1996); hep-ph/9601311.

\bibitem{barger3}V. D. Barger, S. Pakvasa, T. J. Weiler and K. Whisnant, {\it Phys. Rev. Lett.} {\bf 85}, 5055 (2000); hep-ph/0005197.

\bibitem{fardon}R. Fardon, A. E. Nelson and N. Weiner, astro-ph/0309800; 
P. Q. Hung, hep-ph/00010126.

\bibitem{hung}P. Q. Hung and H. Paes, astro-ph/0311131.

\bibitem{hooper}D. Hooper,  D. Morgan and E. Winstanley, hep-ph/0410094.

\bibitem{bilenky1}L. Wolfenstein, {\it Nucl. Phys.} {\bf B186}, 147 (1981);
S. M. Bilenky and B. M. Pontecorvo, {\it Sov. J. Nucl. Phys.} {\bf 38}, 248
(1983); S. T. Petcov, {\it Phys. Lett.} {\bf B110}, 245 (1982).

\bibitem{beacom3}J. F. Beacom, N. Bell, D. Hooper, J. G. Learned, S. Pakvasa 
and T. J. Weiler; {\it Phys. Rev. Lett.}, {\bf 92} (2004); hep-ph/0307151; 
see also P. Keranen, J. Maalampi, M. Myyrylainen and J. Riittinen,
{\it Phys. Lett.} {\bf B574}, 162 (2003); hep-ph/0307041 for similar 
considerations.

\bibitem{kobayashi} M. Kobayashi and C. S. Lim, {\it Phys. Rev.} {\bf D64}, 013003 (2001); hep-ph/0012266.

\bibitem{weiler1}D. J. Wagner and T. J. Weiler, {\it Mod. Phys. Lett.} {\bf A12}, 2497 (1997).

\bibitem{freedman}W. L. Freedman et al. {\it Astrophys. J.} {\bf 553}, 47 (2001).

\bibitem{stodolsky}L. Stodolsky, {\it Phys. Lett.} {\bf B473}, 61 (2000); astro-ph/9911167.

\bibitem{choubey}S. Choubey and S. F. King, {\it Phys. Rev.} {\bf D67}, 073005 (2003), hep-ph/0207260.


\bibitem{beacom4}J. F. Beacom, N. Bell, D. Hooper, S. Pakvasa and T. J.
Weiler; {\it Phys. Rev.} {\bf D68}, 093005 (2003); hep-ph/0307025; F. Halzen and D. Hooper, {\it Rept. Prog. Phys.} 
{\bf 65}, 1025 (2002); astro-ph/0204527.

\bibitem{karle}A. Karle, {\it Nucl. Phys. Proc. Supp.}, {\bf 118} (2003);
astro-ph/0209556; A. Goldschmidt, {\it Nucl. Phys. Proc. Suppl.} {\bf 110},
516 (2002).


\bibitem{barenboin}G. Barenboim and C. Quigg, {\it Phys. Rev.} {\bf D67}, 073024 (2003); hep-ph/0301220.

\bibitem{goldberg}L. A. Anchordoqui, H. Goldberg, F. Halzen and T. J. Weiler,
astro-ph/0311002.
\end{thebibliography}
\end{document}